\documentclass[11pt]{article}
\usepackage{acl2014}
\usepackage{times}
\usepackage{url}
\usepackage{latexsym}
\usepackage{graphicx}
\usepackage{multirow}
\usepackage{subcaption}

\title{Capturing ``attrition intensifying" structural traits from didactic interaction sequences of MOOC learners}

\author{Tanmay Sinha$^1$, Nan Li$^2$, Patrick Jermann$^3$, Pierre Dillenbourg$^2$ \\
  $^1$Language Technologies Institute, Carnegie Mellon University, Pittsburgh PA 15213, USA \\
  $^2$Computer-Human Interaction in Learning and Instruction, EPFL, CH 1015, Switzerland \\
  $^3$Center for Digital Education, EPFL, CH 1015, Switzerland \\
  {\tt  $^1$tanmays@andrew.cmu.edu,  $^2$$^,$$^3$<firstname.lastname>@epfl.ch} \\}

\date{}

\begin{document}
\maketitle
\begin{abstract}
This work is an attempt to discover hidden structural configurations in learning activity sequences of students in Massive Open Online Courses (MOOCs). Leveraging combined representations of video clickstream interactions and forum activities, we seek to fundamentally understand traits that are predictive of decreasing engagement over time. Grounded in the interdisciplinary field of network science, we follow a graph based approach to successfully extract indicators of active and passive MOOC participation that reflect persistence and regularity in the overall interaction footprint. Using these rich educational semantics, we focus on the problem of predicting student attrition, one of the major highlights of MOOC literature in the recent years. Our results indicate an improvement over a baseline ngram based approach in capturing ``attrition intensifying" features from the learning activities that MOOC learners engage in. Implications for some compelling future research are discussed.   
\end{abstract}

\section{Introduction}
Massive Open Online Courses (MOOCs) have attracted millions of students, and yet, their pedagogy is often less elaborated than the state of the art in learning sciences. Scaling up learning activities in MOOCs can be viewed as a sacrifice of pedagogical support, made acceptable by the benefits of giving broad access to education for a marginal increase of costs. Even with students volunteering as teaching assistants in MOOCs, it is not possible to provide at a distance the same support quality in a class of ten thousand as in a class of a hundred, because of the difficulty to collect and analyse data from such a high number of learners. This means that MOOC instructors need to rely on rich computational methods that capture the formalism of how learners progress through the course and what traits of decreasing engagement with the course are predictive of attrition over time. The interpretation of the state of the students can then either be performed by the students themselves, by a human coach or by an automated agent that can deliver recommendations to the students.

In this work, we model the sequence of learning activities in the MOOC as a graph with specific properties. Describing the participants actions sequence as a graph may initially sound as a futile complexity since most MOOCs are built as a simple linear sequence of activities (watch video, do assignments, read forums). However, when looking at the activity in more detail, some sequences are richer and justify a more powerful descriptive modeling. The descriptive power of the graph model is to capture the underlying structure of the learning activity. The hypothesis is that formalizing the workflow of such heterogeneous behavior in MOOCs, is one solution to be able to a) scale up learning activities that may initially appear as non scalable, b) help instructors reason out how educational scenarios concretely unfold with time, such as what happened during the course (at what times were learners active and performing well, lost, disoriented or trapped) and what needs to be repaired.

\section{Related Work}
In this section we outline perspectives on student attrition that have been explored so far in the literature on MOOCs. Much of this work successfully leverages effective feature engineering and advanced statistical methods. However, the biggest limitation of most of these emerging works is that they focus solely on discussion forum behavior or video lecture activity, but do not fuse and take them into account. Some of these works have grown out of research on predicting academic progress of students and identifying students those who are at dropout risk \cite{Kot:03,Dek:09,Pal:12,Mar:13,Man:14}.

Some prior research has focused on deriving social positioning metrics within discussion forums to understand influencing factors that lead to differently motivated behaviors of students. For example, \cite{Yang:13,Rose:14} used aggregate post-reply discussion forum graph per week, with an aim to investigate posting behavior and collaborative aspects of participation through operationalizations of social positioning. However, we work at a much finer granularity in the current study and our focus is on individual student modeling instead. We capture not only forum participation trajectory, but also video lecture viewing activity of every student in their participation week. Modeling the combined interaction footprint as an activity network, allows us to decipher the type of engagement and organization of behavior for each student, which are reflective of attrition. 

Similarly \cite{Ram:14,WenA:14,WenB:14} published results that describe longitudinal discussion forum behavior affecting student dropout, in terms of posting, viewing, voting activity, level of subjectivity (cognitive engagement) and positivity (sentiment) in students' posts. Related to this, one recent work of \cite{Rossi:14} have made an attempt to overcome the language dependency drawback of these works and capture language independent discussion forum features related to structure, popularity, temporal dynamics of threads and diversity of students.

It is important to note, however, that all this substantial research caters to only about 5\% of students who participate in MOOC discussion forums \cite{Huang:14}. Our recent work has laid a preliminary foundation for research investigating students' information processing behavior while interacting with MOOC video lectures \cite{SinhaA:14}. We apply a cognitive video watching model to explain the dynamic process of cognition involved in MOOC video clickstream interaction and develop a simple, yet potent information processing index that can be effectively used as an operationalization for making predictions regarding critical learner behavior, specifically in-video and course dropouts. In an attempt to better understand what features are predictive of students ceasing to actively participate in the MOOC, \cite{Veer:14} have integrated a crowd sourcing approach for effective feature engineering at scale. Among posting, assignment and grading metrics, students' cohort membership depending on their MOOC engagement was identified as an influential feature for dropout prediction.

\section{Study Context}
The current study is a part of the shared task for EMNLP 2014 Workshop on Modeling Large Scale Social Interaction in Massively Open Online Courses \cite{RoseA:14}. We have both video clickstream data (JSON) and discussion forum activity data (SQL) from one Coursera MOOC as training data, that we use in this work. Our predictive models will also be tested on 5 other Coursera MOOCs. 

In general, Coursera forums, divided into various subforums, have a thread starter post that serves as a prompt for discussion. The thread builds up as people start following up discussions by their posts and comments. As far as our forum dataset is concerned, we have 31532 instances of forum viewing and 35306 instances of thread viewing. In addition to this view data, we have 4840 posts and 2652 comments among 1393 threads initiated in the discussion forums during the span of the course, which received 5060 upvotes and 1763 downvotes in total.

To supplement the forum data, we additionally leverage rich video interaction data from the clickstream data. The clickstream data contains many errors. We obtained 82 unique video ids from the clickstream data, but only 45 of them are valid (watched by large number of unique students). The 37 invalid video ids may be simply due to logging errors. They are also likely to be videos that were uploaded by the course staff for testing purposes. There are in total 27739 students registered the course, however, only 14312 students had online video interactions. The rest of the students may have never logged in, or only have viewed the course pages, or have downloaded the videos without further online engagement. Among the 14312 students who have video interactions, 14264 of them have valid video events logged, which lead to 181100 valid video sessions for our analysis. These valid video sessions further contain 462341 play events, 295103 pause events, 87585 forward jumps, 98169 backward jumps, 6707 forward scrolls, 5311 backward scrolls, 18051 video-play rate increase and 16163 decrease events, respectively. 

Our dropout prediction approach that will be described in the next section is applied to student interactions comprising of only online forum and video viewing activities. Currently, we do not make use of the pageview click data.

% COMMENT I think we can remove "firstly" "secondly" etc. since enumerate prints numbers already.
\section{Technical Approach}

\begin{enumerate}
\itemsep-0.2em
\item To capture the behaviors exhibited in two primary MOOC activities, namely video lecture viewing and forum interaction, we operationalize the following metrics:
\begin{itemize}
\item {\bf Video lecture clickstream activities}: Play (PL), Pause (PA), SeekFw (FW), SeekBw (BW), ScrollFw (FS), ScrollBw (BS), Ratechange Increase (RCI), Ratechange Decrease (RCD). When two seek events happen in $<1$ second, we group them into a scroll. We encode ratechange event based on whether students sped up or slowed down with respect to playrate of the last click event. 
\item {\bf Discussion forum activities}: Post (Po), Comment (Co), Thread (Th), Upvote (Uv), Downvote (Dv), Viewforum (Vf), Viewthread (Vt) 
\end{itemize}

\item Because timing of all such MOOC events are logged in our data, we sort all these activities by timestamp to obtain the sequence of activities done by students. This gives us a simple sequentially ordered time series that can be used to reason about behavioral pattern of students.

\item We form the interaction footprint sequence for students by concatenating all their different timestamped MOOC activities for every week of MOOC activity. For example, if a student watched a video (PL, PA, FW, RCI, PA) at [time {\em `i'}, week {\em `j'}], viewed a forum at time [{\em `i+1'}, week {\em `j'}] and consequently made a post at [time {\em `i+2'}, week {\em `j'}], his interaction footprint sequence for week `j' would be: PL PA FW RCI PA Vf Po. Forming such a sequence captures in some essence, the cognitive mind state that govern students' interaction, as they progress through the MOOC by engaging with these multiple forms of computer mediated inputs. Most MOOCs are based on a weekly rhythm with a new set of videos and new assignments released every week.  

\item To find subsequences that might help us to predict student dropout before it occurs, we extract the following set of features for each student in each of his participation weeks:
\begin{itemize}
\itemsep-0.2em
\item {\bf N-grams} from the interaction footprint sequence (n = 2 to 5). Such `n' consecutively occurring MOOC activities not only characterize suspicious behaviors that might lead to student attrition but also help us to automatically determine the elements of what might be considered ``best MOOC interaction practices" that keep students engaged.
\item {\bf Proportion} of video viewing activities among all video interactions, that are active or passive. We define passive video viewing as mere play and pause (PL, PA), while rest of the video lecture clickstream activities (FW, BW, FS, BS, RCD, RCI) are considered elements of active video viewing.
\item {\bf Proportion} of discussion forum activities among all forum interactions, that are active or passive. We define passive forum activities as viewing a forum or thread (Vf, Vt), upvoting and downvoting (Uv, Dv). The forum activities of starting a thread (Th), posting (Po) and commenting (Co) are indicative of active forum interaction.
\end{itemize}
In general, because passive video lecture viewing is high (for example, 48\% of all video clickstream activities in our dataset comprise of activity sequences having only PL event), discussion forum conversation networks in MOOCs are sparse (only 10\% of forum activities relate to explicitly posting, commenting or starting a thread) and passive forum activities are very predominant (90\% of forum interactions in our dataset are just passively viewing a thread/forum, upvoting or downvoting), differentiating between such active and passive forms of involvement might clarify participation profiles that are most likely to lead to disengagement of students from the MOOC.

\item In an attempt to enrich the basic ngram representation and better infer traits of active and passive participation, we extract the following set of graph metrics from the overall interaction footprint sequence. Specifically, in this modeling scheme, we extract consecutive windows of length two and create a directed edge of weight one between the activities appearing in sequential order. This results in a directed graph (having self loops and parallel edges), with nodes representing activities done by a student in particular week, while the weighted edges representing the frequencies of activities appearing after one another. For example, in a sequence, (Vt Po Vt Po Po), corresponding nodes in the graph are Vt and Po, while edges are (Vt, Po), (Po, Vt), (Vt, Po) and (Po, Po). The activity graph thus describes the visible part of the educational activities (who does what and when) and models the structure of activity sequences, rather than the details of each activity. Features from the syntactic structure of the graph along with their educational semantics are described below.
\begin{itemize}
\itemsep-0.1em
\item {\bf Number of nodes and edges}: Indicative of whether overall participation of students in different MOOC activities is high or low.
\item {\bf Density}: Graph density is a tight-knittedness indicator of how involved students are in different MOOC activities, how clustered their activities are or how frequently they switch back and forth between different activities. Technically, for a directed network, density = $m/n(n-1)$, where m=number of edges, n=number of nodes. For our multidigraph representation, density can be $>$1, because self loops are counted in the total number of edges. This also implies that values of density $>$1 denote high persistence in doing particular set of MOOC activities, because of greater number of self loops. 
\item {\bf Number of self loops}: Though graph density provides meaningful interpretations when $>1$, we can't conclusively infer activity persistence in an activity graph with low density. So, we additionally extract number of self loops to refer to the regularity in interaction behavior.
\item {\bf Number of Strongly Connected Components (SCC)}: SCC define a special relationship among a set of graph vertices that can be exploited (each vertex can be reached from every other vertex in the component via a directed path). If the number of SCC in an activity graph are high, there is a high probability that students performs certain set of activities frequently to successfully achieve their desired learning outcomes in the course. This might be an influential indicator for behavioral organization and continuity reflected in overall interaction footprint of students. Dense networks are more likely to have greater number of SCC.
\item {\bf Central activity}: We extract top three activities of students with maximum indegree centrality, for each of their participation weeks. Technically, indegree centrality for a node {\em `v'} is the fraction of nodes its incoming edges are connected to. Depending on which are the central activities of students, we can characterize how active or passive is the participation. For example, Viewthread and Viewforum (Vt, Vf) are more passive forms of participation than Upvote and Downvote (Uv, Dv), which are in turn more passive than Posting, Commenting, Thread starting (Po, Co, Th) and other intense forms of video lecture participation that represent high grappling with the course material.
\begin{figure*}[t]
\centering
        \begin{subfigure}[b]{0.22\textwidth}
                \includegraphics[width=\textwidth]{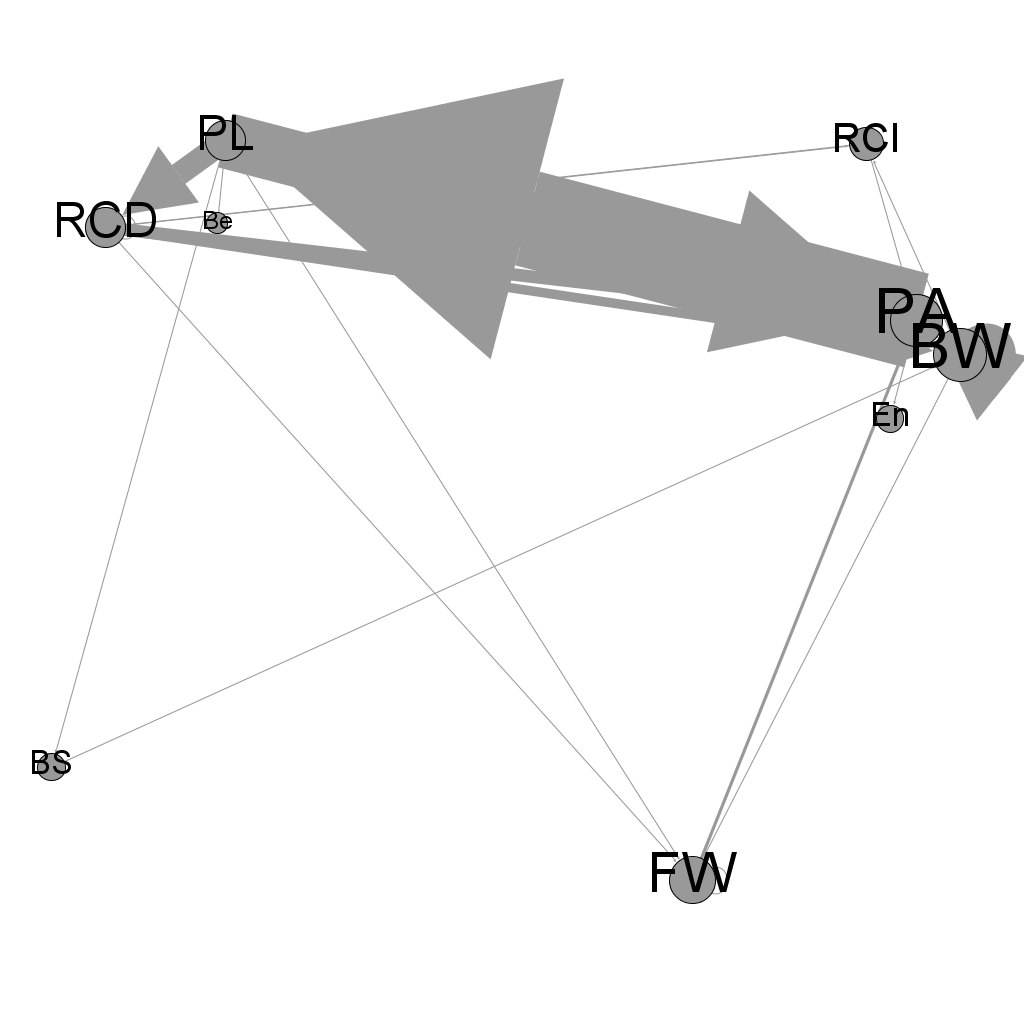}
                \caption{Active video viewing}
                \label{fig:gull}
        \end{subfigure}%
\quad
        ~ %add desired spacing between images, e. g. ~, \quad, \qquad, \hfill etc.
          %(or a blank line to force the subfigure onto a new line)
        \begin{subfigure}[b]{0.22\textwidth}
                \includegraphics[width=\textwidth]{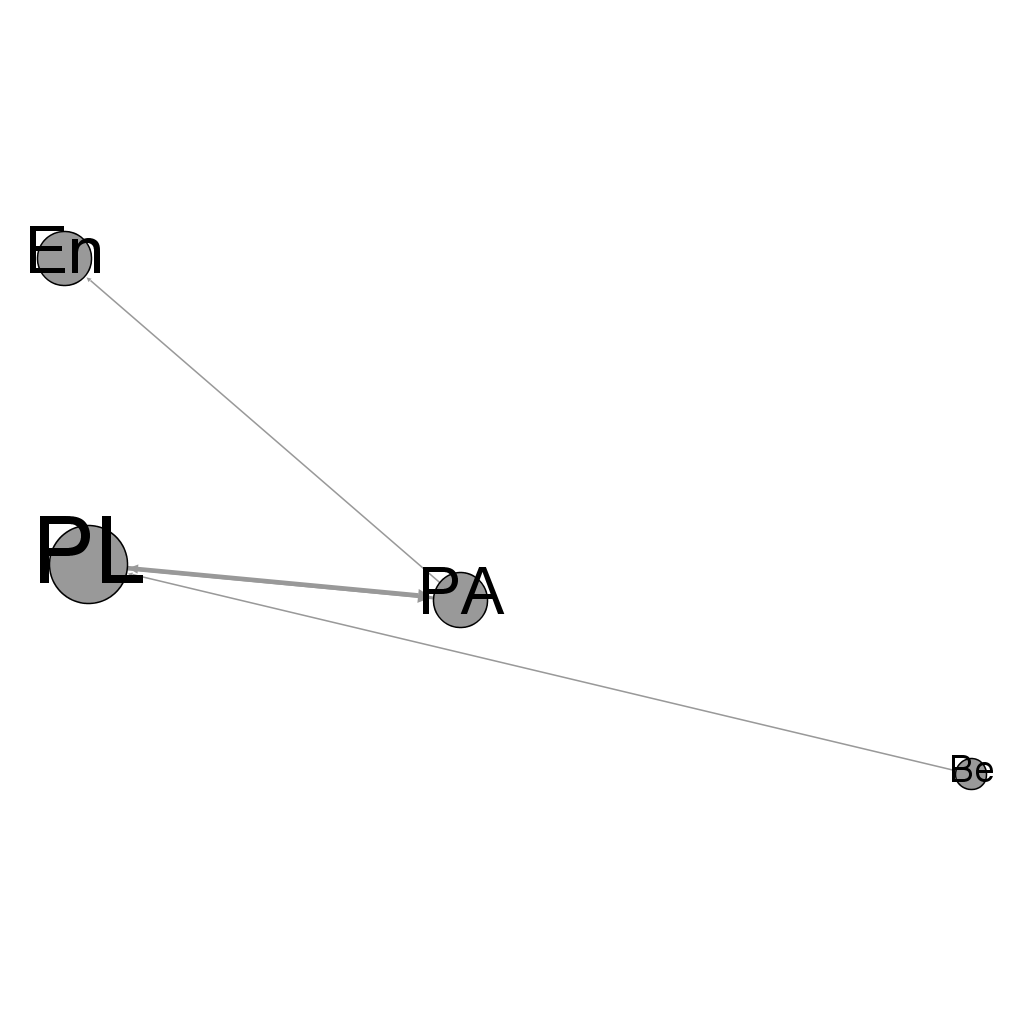}
                \caption{Passive video viewing}
                \label{fig:tiger}
        \end{subfigure}
\quad
        \begin{subfigure}[b]{0.22\textwidth}
                \includegraphics[width=\textwidth]{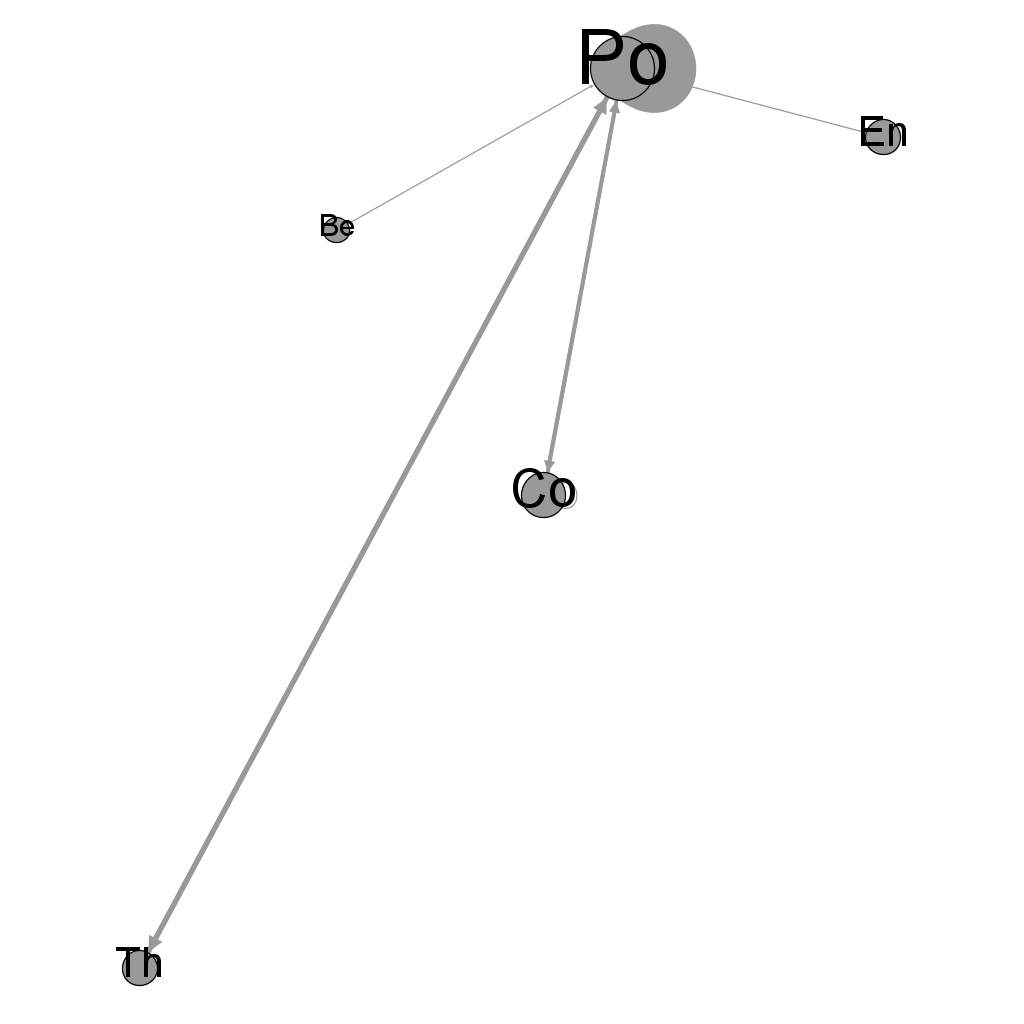}
                \caption{Active forum activity}
                \label{fig:tiger}
        \end{subfigure}
\quad
        \begin{subfigure}[b]{0.22\textwidth}
                \includegraphics[width=\textwidth]{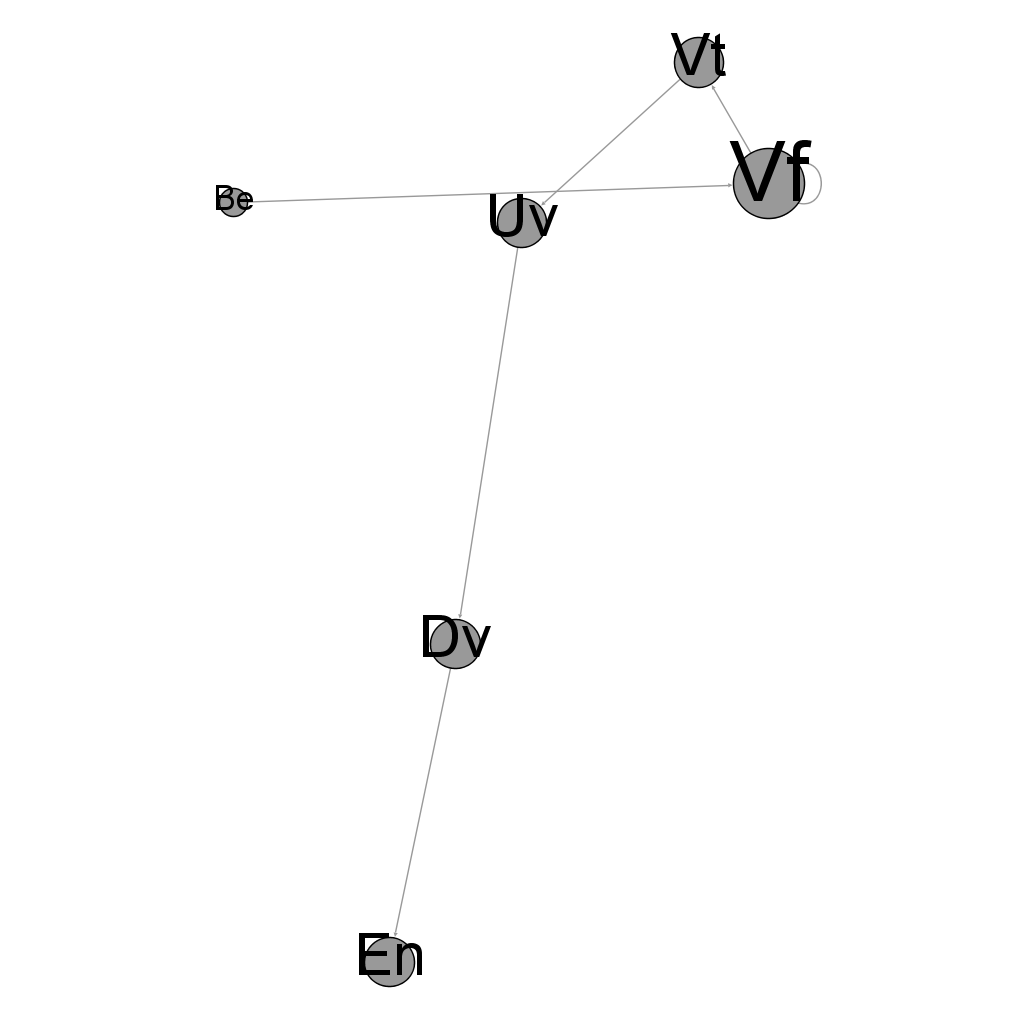}
                \caption{Passive forum activity}
                \label{fig:tiger}
        \end{subfigure}
\caption{Interaction graphs representing 4 contrasting MOOC scenarios in our dataset}
\label{fig:1}
\end{figure*}
\item {\bf Central transition}: We extract the edge (activity transition) with maximum betweenness centrality, which acts like a facilitator in sustaining or decreasing participation. Technically, betweenness centrality of an edge {\em `e'} is the sum of the fraction of all-pairs shortest paths that pass through {\em `e'}. We normalize by $1/n(n-1)$ for our directed graphical representation, where {\em `n'} is the number of nodes. For example, Vt-Po (view thread-post) could be one of the central edges for Th (thread starting activity), which in turn is a strong student participation indicator. Alternately, Po/Co/Th-Dv (post/comment/thread initiate-downvote) could serve as decision conduits that increase dissatisfaction of students because of others' off content/off-conduct posting. Such lack of exposure to useful and informative posts on forums can potentially aggravate feelings of ``lack of peer support" and ``healthy community involvement", inturn leading to decreasing engagement.
\end{itemize}
\item We add certain control variables in our feature set to account for inherently present student characteristics, namely {\bf courseweek} (number of weeks since the course has been running), {\bf userweek} (number of weeks since the student joined the course) and a {\bf nominal variable} indicating whether student activity in a week comprised of only video lecture viewing, only forum activity, both or none.
\end{enumerate}
Because we are interested in investigating a)how behavior within a week affects students' dropout in the next course week, b)how cumulative behavior exhibited up till a week affects students' dropout in the next course week, we create two experimental setups: one using data from the current participation week ({\bf{\em Curr}}) and the second using data from the beginning participation week till the current week ({\bf{\em TCurr}}). For the second setup, all feature engineering is done from the cumulative interaction footprint sequence. 

Some of the interaction graphs culled out from the footprint sequence, which are representative of active and passive MOOC participation are depicted in figure 1. Each graph has a begin (Be) and end (En) node, with nodes sized by indegree centrality and directed edges sized by tie strength.

\section{Results}
\subsection{Evaluating Our Features}
As we would intuitively expect, mean and standard deviations for all our extracted graph metrics are higher in the {\bf{\em TCurr}} setup. Another evident pattern is that all these graph metrics follow long tailed distributions for both {\bf{\em Curr}} and {\bf{\em TCurr}} setups, with very few students exhibiting high values. These distributions concur with the 90-9-1 rule in online communities which says that 90\% of the participants only view content (for example, watch video, Vf, Vt), 9\% of the participants edit content (for example, Uv, Dv), and 1\% of the participants actively create new content (for example, Po, Co, Th). Moreover, we notice that the top three central activities with maximum frequency and central edges that describe interactions between them, are passive interaction events. Among the top 20, we can observe central edges such as RCI-RCI or PL-FW that hint towards skipping video and hence decreasing participation, while Th-PL, Po-PL, Th-Po that point towards facilitating participation. Thus, in order to graphically visualize interactions among features and their relationship to the class distribution (dropout and non dropout), we utilize mosaic plot representation. The motivating question being two-fold: a)How do the extracted features vary among dropouts and non dropouts?  b)When viewing more than one features together, what can we say about association of different feature combinations to survival of students in the MOOC? After ranking feature projections on basis of interaction gain (in \% of class entropy removed), we discern the following:
\begin{table*}[t]
\centering
\begin{tabular}{|p{3.5 cm}|p{3.5 cm}|p{2.2 cm}|p{2.4 cm}|}
\hline \bf Model & \bf Performance Metric & \bf Setup {\bf{\em Curr}} & \bf Setup {\bf{\em TCurr}} \\ \hline
1. Baseline  & \multirow{2}{*} {}Accuracy/Kappa & 0.623/0.297 & 0.647/0.173 \\  
&  False Negative Rate  & 0.095 & 0.485 \\ \hline
2. Graph  & \multirow{2}{*} {}Accuracy/Kappa & 0.692/0.365 \# & 0.693/0.277 \#\\  
& False Negative Rate & 0.157 & 0.397 \\ \hline
3. Baseline + Graph  & \multirow{2}{*} {}Accuracy/Kappa & 0.624/0.298 & 0.646/0.173\\  
& False Negative Rate & 0.095 & 0.482 \\ \hline 
\end{tabular}
\caption{\label{FZM} Performance metrics for machine learning experiments. Random classifier performance is 0.5. Values marked \# are significantly better (p$<$0.01, pairwise t-test) than other results in same column}
\end{table*}

\begin{itemize}
\itemsep-0.3em
% COMMENT why is the second bullet mixing joining week and note attributes. Wouldn't it be clearer to include SCC and numer of nodes in the first bullet ? 
\item For both {\bf{\em Curr}} and {\bf{\em TCurr}} setups, the mosaic plots reveal that dropout is higher for students having low number of nodes, edges, SCC and self loops, low activity graph density, low proportion of active forum and video viewing activity. This reflects that our operationalizations drawn from overall interaction footprint are successfully able to capture features expressing student behavior that might escalate attrition.
\item Student dropout is higher if they join in later course weeks and have a sparse activity graph. There could be 2 possible explanations: a)Students join later and do minimal activity because they only have specific information needs. So, they do not stay after interacting with the course material in a short non linear fashion and satisfying their needs, b)Students who join later are overwhelmed with lots of introductory and prerequisite MOOC video lectures to watch, pending assignments to be completed to successfully pass the course and discussion forum content already posted. Finding difficulty in coping up with the ongoing pace of the MOOC, they do not stay for prolonged periods in the course.
\end{itemize} 
\subsection{Dropout Prediction and Analysis}
We leverage machine learning techniques to predict student attrition along the way based on our extracted feature set. The dependent class variable is dropout, which is 0 for all active student participation weeks and 1 only for the last participation week (student ceased to participate in the MOOC after that week), leading to an extremely skewed class distribution. Note that by active student participation, we refer to only forum and video viewing interactions. We construct the following two models for validation. For each model, there is a {\bf{\em Curr}} and a {\bf{\em TCurr}} setup:
\begin{itemize}
\itemsep-0.3em
\item {\bf Baseline Ngram Model}: Features used are Coursweek, Userweek, Ngrams from full interaction footprint sequence (2 to 5), Ngram length, proportion of active/passive video viewing and forum activity (dichotomized by equal width), nominal variable. 
\item {\bf Graph Model}: Features used are Coursweek, Userweek, Ngram length, Graph metrics (top 3 central activities, density (dichotomized by equal frequency), central transition, no. of nodes (dichotomized by equal frequency), no. of edges (dichotomized by equal frequency), no. of self loops (dichotomized by equal frequency), no. of SCC), nominal variable. 
\end{itemize}
For both these models, we use cost sensitive LibSVM with radial basis kernel function (RBF) as the learning algorithm \cite{Hsu:03}. The advantage of RBF is that it nonlinearly maps samples into a higher dimensional space so it, unlike the linear kernel, can handle the case when the relation between class labels and attributes is nonlinear. Rare threshold for feature extraction is set to 4, while cross validation is done using a supplied test set with held out students having sql id 798619 through 1882807.

The important take away messages from these results are:

\begin{itemize}
\item Graph model performs significantly better than Baseline ngram model for both {\bf{\em Curr}} (t=-17.903, p$<$0.01) and {\bf{\em TCurr}} (t=-11.834, p$<$0.01) setups, in terms of higher accuracy/kappa and comparable false negative rates\footnote{False negative rate of 0.x means that we correctly identify (100-(100*0.x))\% of dropouts}. This is because the graph models the integration of heterogeneous MOOC activities into a structured activity. The edges of the graph, which connect consecutive activities represent a two-fold relationship between these activities: how they relate to each other from a pedagogical and from an operational viewpoint. In addition to capturing just the order and mere presence of active and passive MOOC events scatterred throughout the activity sequence, the activity network representation additionally captures different properties of MOOC interaction such as a)how recurring behaviors develop in the participation trajectory of students, and how the most central ones thrust towards increasing or decreasing engagement, b)how the number and distribution of such activities are indicative of persistence in interaction behavior. The baseline+graph approach does not lead to improvement in results over the baseline approach.
\item {\bf{\em TCurr}} setup does not necessarily lead to better results than {\bf{\em Curr}} setup. This indicates that students' attrition is more strongly influenced by the most recent week's exhibited behavioral patterns, rather than aggregated MOOC interactions from the beginning of participation. The extremely small false negative rates in {\bf{\em Curr}} setup indicate the effectiveness of our feature engineering approach in predicting attririon behavior, even with an extremely skewed class distribution. However, more studies would be required to corroborate the relation between change in interaction sequences from one week to another and factors such as students' confusion (``I am unable to follow the course video lectures") or negative exposure (``I am not motivated enough to engage because of less productive discussion forums"), which gradually build up like negative waves before dropout happens \cite{SinhaB:14}. 
\end{itemize} 

\section{Conclusion and Future Work}
In this work, we formed operationalizations that quantify active and passive participation exhibited by students in video lecture viewing and discussion forum behavior. We were successful in developing meaningful indicators of overall interaction footprint that suggest systematization and continuity in behavior, which are in turn predictive of student attrition. In our work going forward, we seek to differentiate the interaction footprint sequences further using potent markov clustering based approaches. The underlying motivation is to decipher sequences having lot of activity overlap as well as similar transition probabilities. These cluster assignments can then serve as features that help segregating interaction sequences predictive of dropout versus non-dropouts. 

Another interesting enhancement to our work would include grouping commonly occurring activities that learners perform in conjunction with each other and form higher level latent categories indicative of different participation traits. In our computational work, we have recently been developing techniques for operationalizing video lecture clickstreams of students into cognitively plausible higher level behaviors to aid instructors to better understand MOOC hurdles and reason about unsatisfactory learning outcomes \cite{SinhaA:14}.

One limitation of the above work is that we are concerned merely with the timestamped order of activities done by a student and not the time gap between activities appearing in the interaction footprint sequence. The effect of an activity on a subsequent activity often fades out with time, i.e. as the lag between two activities increases: learners forget what they learned in a previous activity. For example, the motivation created at the beginning of a lesson by presenting an interesting application example does not last forever, so as to initiate productive forum discussions. Similarly, the situation of a thread being started (Th) and a post being made (Po) within 60 secs of completing video lecture viewing, might imply a different behavior, than if these forum activities occur five days after video lecture viewing. Therefore, we seek to better understand context of the most and least central activities of students in MOOCs, differentiating between subsequences lying within and outside user specified temporal windows. Our goal is to view the interaction footprint sequence formation in a sequential data mining perspective \cite{Moon:13} and discover a)most frequently occurring interaction pathways that lead students to such central activities, b)association rules with high statistical confidences that help MOOC instructors to trace why students engage in certain MOOC activities. For example, a rule of the form AB $\Rightarrow$ C, such as ``Vf", ``Uv" [15s] $\Rightarrow$ ``Po" [30s] (confidence = 0.7), is read as if a student navigated and viewed a forum page followed by doing an upvote within 15 seconds, then within the next 30 seconds he would make a post 70\% of the time.

% include your own bib file like this:
%\bibliographystyle{acl}
%\bibliography{acl2014}

\begin{thebibliography}{}

\bibitem[\protect\citename{Dekker et al.}2009]{Dek:09}
Dekker, G. W., Pechenizkiy, M., \& Vleeshouwers, J. M. (2009). ``Predicting Students Drop Out: A Case Study". {\em International Working Group on Educational Data Mining}.

\bibitem[\protect\citename{Huang et al.}2014]{Huang:14}
Huang, J., Dasgupta, A., Ghosh, A., Manning, J., and Sanders, M. 2014. ``Superposter behavior in MOOC forums". {\em ACM Learing at Scale(L@S)}

\bibitem[\protect\citename{Hsu et al.}2003]{Hsu:03}
Hsu, C. W., Chang, C. C., \& Lin, C. J. (2003). ``A practical guide to support vector classification"

\bibitem[\protect\citename{Kotsiantis et al.}2003]{Kot:03}
Kotsiantis, S. B., Pierrakeas, C. J., \& Pintelas, P. E. (2003, January). ``Preventing student dropout in distance learning using machine learning techniques". {\em In Knowledge-Based Intelligent Information and Engineering Systems} (pp. 267-274). Springer Berlin Heidelberg.

\bibitem[\protect\citename{Manhaes et al.}2014]{Man:14}
Manhaes, L. M. B., da Cruz, S. M. S., \& Zimbrao, G. (2014, March). ``WAVE: an architecture for predicting dropout in undergraduate courses using EDM". {\em In Proceedings of the 29th Annual ACM Symposium on Applied Computing} (pp. 243-247). ACM.

\bibitem[\protect\citename{M\'arquez-Vera et al.}2013]{Mar:13}
M\'arquez-Vera, C., Cano, A., Romero, C., \& Ventura, S. (2013). ``Predicting student failure at school using genetic programming and different data mining approaches with high dimensional and imbalanced data". {\em Applied intelligence}, 38(3), 315-330.

\bibitem[\protect\citename{Mooney and Roddick}2013]{Moon:13}
Mooney, C. H., \& Roddick, J. F. (2013). ``Sequential pattern mining--approaches and algorithms". {\em ACM Computing Surveys (CSUR)}, 45(2), 19.

\bibitem[\protect\citename{Pal}2012]{Pal:12}
Pal, S. (2012). ``Mining educational data to reduce dropout rates of engineering students". {\em International Journal of Information Engineering and Electronic Business (IJIEEB)}, 4(2), 1.

\bibitem[\protect\citename{Ramesh et al.}2014]{Ram:14}
Ramesh, A., Goldwasser, D., Huang, B., Daume III, H., \& Getoor, L. (2014, June). ``Learning latent engagement patterns of students in online courses". {\em In Twenty-Eighth AAAI Conference on Artificial Intelligence}.

\bibitem[\protect\citename{Ros\'e et al.}2014]{Rose:14}
Ros\'e, C. P., Carlson, R., Yang, D., Wen, M., Resnick, L., Goldman, P., \& Sherer, J. (2014, March). ``Social factors that contribute to attrition in moocs. {\em In Proceedings of the first ACM conference on Learning@ scale conference} (pp. 197-198). ACM.

\bibitem[\protect\citename{Ros\'e and Siemens}2014]{RoseA:14}
Ros\'e, C. P., Siemens, G. (2014).  ``Shared Task on Prediction of Dropout Over Time in Massively Open Online Courses", {\em Proceedings of the 2014 Empirical Methods in Natural Language Processing Workshop on Modeling Large Scale Social Interaction in Massively Open Online Courses}, Qatar, October, 2014.

\bibitem[\protect\citename{Rossi and Gnawali}2014]{Rossi:14}
Rossi, L. A., \& Gnawali, O. ``Language Independent Analysis and Classification of Discussion Threads in Coursera MOOC Forums".

\bibitem[\protect\citename{Sinha et al.}2014]{SinhaA:14}
Sinha, T., Jermann, P., Li, N., Dillenbourg, P. (2014). ``Your click decides your fate: Inferring Information Processing and
Attrition Behavior from MOOC Video Clickstream Interactions". {\em Proceedings of the 2014 Empirical Methods in Natural Language Processing Workshop on Modeling Large Scale Social Interaction in Massively Open Online Courses}, Qatar, October, 2014.

\bibitem[\protect\citename{Sinha}2014]{SinhaB:14}
Sinha, T. (2014). ``Who negatively influences me? Formalizing diffusion dynamics of negative exposure leading to student attrition in MOOCs". {\em LTI Student Research Symposium, Carnegie Mellon University}

\bibitem[\protect\citename{Veeramachaneni et al.}2014]{Veer:14}
Veeramachaneni, K., O'Reilly, U. M., \& Taylor, C. (2014). ``Towards Feature Engineering at Scale for Data from Massive Open Online Courses". {\em arXiv preprint arXiv:1407.5238}.

\bibitem[\protect\citename{Wen et al.}2014a]{WenA:14}
Wen, M., Yang, D., \& Ros\'e, C. P. (2014a). ``Linguistic Reflections of Student Engagement in Massive Open Online Courses". {\em In Proceedings of the International Conference on 	Weblogs and Social Media} 

\bibitem[\protect\citename{Wen et al.}2014b]{WenB:14}
Wen, M., Yang, D., \& Ros\'e, C. P. (2014b). ``Sentiment Analysis in MOOC Discussion Forums: What does it tell us?". {\em In Proceedings of Educational Data Mining}

\bibitem[\protect\citename{Yang et al.}2013]{Yang:13}
Yang, D., Sinha T., Adamson D., and Rose, C. P. 2013. ``Turn on, Tune in, Drop out: Anticipating student dropouts in Massive Open Online Courses" {\em In NIPS Workshop on Data Driven Education}
\end{thebibliography}

\end{document}